\documentclass[floatfix,twocolumn,rmp]{revtex4}

\usepackage{dcolumn,graphicx,amsmath,amssymb}
\usepackage{txfonts}

\begin{document}

\title{Modelling the dynamics of youth subcultures}

\author{Petter Holme}
\affiliation{Department of Physics, University of Michigan, Ann Arbor,
  MI 48109}

\author{Andreas Gr\"{o}nlund}
\affiliation{Department of Physics, Ume{\aa} University, 90187
  Ume{\aa}, Sweden}

\begin{abstract}
What are the dynamics behind youth subcultures such as punk, hippie,
or hip-hop cultures? How does the global dynamics of these subcultures
relate to the individual's search for a personal identity? We propose
a simple dynamical model to address these questions and find that only
a few assumptions of the individual's behaviour are necessary to
regenerate known features of youth culture.
\end{abstract}

\maketitle

\section{Introduction}

One can often distinguish youths from adults, not only by their faces,
but also by their jargon, clothing, gait and posture. Their relative lack of
experience and generally different position in life have probably
always separated the youth and adult worlds~\cite{brake}. The prosperity
of Western societies during the 20th century has postponed the entry
of adulthood. This has increased the importance of youth culture in
the search of identity. It has also diversified youth culture so that
one (in academia since ref.~\cite{rtr}) speaks of youth
subcultures rather than one youth culture. Whereas ``subculture,'' in
general, refers to a social group with particular behaviours or beliefs,
this paper specifically focuses on youth subcultures. Today, youth
subcultures are conspicuous features of Western and other societies of
the world. We note that youth subcultures, as known in the
West, often require some economic strength from the follower. Many aspects
of our discussion are probably ubiquitous, but to simplify we focus on
youth culture in wealthy democracies. Subcultures may be long lived like
the hip hop or punk cultures, or die out almost as soon as they have a
name. They may be centred on music, sports (such as the surf or
snowboard culture~\cite{2ext}) or literature (such as
the Beat generation~\cite{polsky}), religion~\cite{damrell}, etc. The
reason a subculture may flare up and vanish soon is related to that
adolescents in search of an identity are very adaptive in terms of the
defining traits of a subculture. An adolescent's formation of a
personality is both an individual and a social
process~\cite{conger,erksn}. For the social part the subcultures
often play a  major role. In this study we attempt to model the
dynamics of subcultures based on a few assumptions about the rationale
behind the subcultures' roles in the formation of a personal
identity. Our goal is to observe the skewed distribution of sizes of
the subcultures---like Thornton's example about the club subcultures
of late-eighties Britain: ``Hundreds of dance genres are coined every
year. While most fail, acid house prospered.'' (Ref.~\cite{thornton},
p.~156.) ``Genre'' refers to a division of an art form, in this case
club music. Just as music can be classified hierarchically, one can
divide subcultures into sub-subcultures. Since, in the club music
world of Thornton, the subcultural identity is closely tied to the
music, it is natural to associate the genre with a subculture. This is,
of course, a very fine distinction of a subculture; henceforth, we
will not explicitly state the level of the hierarchy we are
discussing, but presume that the same precepts govern subcultures at
all different hierarchical levels. To model youth subcultures is, no
doubt, an exceedingly complex problem; and any approach runs the risk
of either being an over-simplification or so complex the actual
mechanisms become indecipherable. Since this question is, to our
knowledge, not earlier addressed, we propose a simple framework that
can easily be extended to more elaborate future models. We mention
that this model can be applied to different social and economic
systems. In this paper, however, we focus solely on the question of
youth subcultures.

\section{Dynamical model}

Our model is based on five major assumptions: \textit{1.} The dynamics
of the underlying social network is negligible. This is probably our
least realistic assumption---as a youth takes up a new subculture she,
or he, will likely meet and get acquainted with other aficionados. On
the other hand, there are many factors that inhibit the dynamics of
one's social network: the classmates, neighbours and relatives do not
change that rapidly. We anticipate future models that will include an
adaptive network dynamics. For this work we will use static underlying
networks. Some evidence that youth retains some part of their social network
after the entry into adulthood can be found in
Ref.~\cite{degenne:adult}. \textit{2.} An adolescent belongs to one
subculture at a time. This is natural since, belonging to many
subcultures at the same time would conflict with the subculture as a
basis for a personal identity. Furthermore, for many subcultures, a
member is expected to be devoted to it. That much said, it is
sometimes more appropriate to represent the identities in a
multi-dimensional space~\cite{jmmp:blau}. Certainly our study can be
extended to incorporate this framework (someone can be a hip-hopper in
a music dimension and a basketball player in a sports dimension). Note
that, if the dimensions are weakly coupled (interrelated), each
dimension can be treated as a separate, one-dimensional system. For
the present study we assume one-dimensional identities. \textit{3.} If
the fraction of friends that have adopted a certain subculture is big
enough then an adolescent will adopt that subculture too. This
assumption lies behind much of modelling for the spread of
fads~\cite{watts:fad}. It is known that friends play a great role for
an individual's adoption of subcultures~\cite{conger,erksn}. If many
friends of a youth follow a new fad, naturally she, or he, will be
interested in the new thing and may also feel left behind. \textit{4.}
The attractiveness of a subculture decreases with its age. Youths do
not want to be old-fashioned, so a subculture on its decline certainly
seems less appealing to potential followers than a subculture on its
rise. \textit{5.} There is a certain resistance to changing
subcultures. To take up a subculture requires an effort---one needs to
learn the unstated rules and most likely buy various
paraphernalia---i.e.\ acquire subcultural
capital~\cite{thornton}. Then, of course, the followers of a
subculture like what they are doing, if they are not tired and
dissatisfied with what they are doing they would not change.

\begin{figure}
  \resizebox*{\linewidth}{!}{\includegraphics{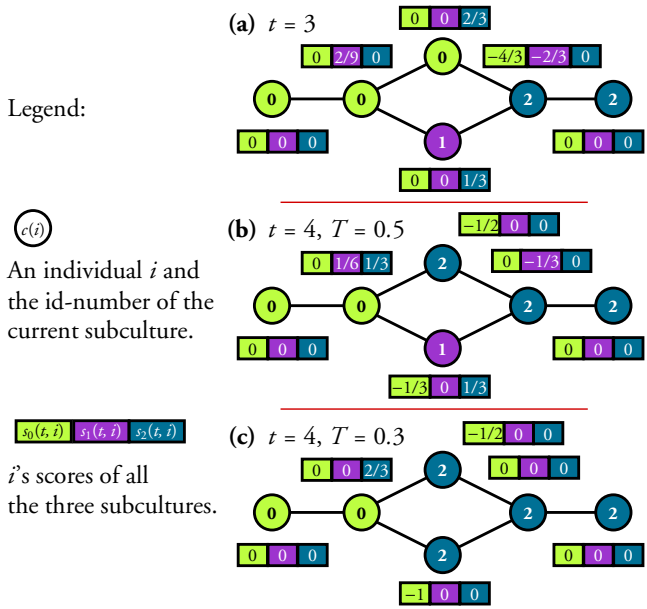}}
  \caption{
    Illustration of the model on a small example graph. The colour and
    number in the circles indicates the current subculture of the
    vertex. The bar next to the circle shows the vertex' score-function
    values for the three subcultures (0, 1, and 2) the current time $t$. The
    starting time of a subculture $t(i)$ is, in this particular example,
    identical to $i$. (b) shows the time-step after (a) in with the
    threshold value $T = 0.5$. In this case no other subculture is
    introduced, no vertex will change subculture as time tends to
    infinity. (c) is the corresponding plot with $T = 0.3$. In this case the
    system will, as time progress, reach a state where every vertex adopts
    subculture 2.
  }
  \label{fig:ill}
\end{figure}

Many subcultures define themselves against the ``mainstream''---the
commercialised culture promoted by media~\cite{thornton,brake}. It has
been argued that this dichotomy---the subcultures vs.\ the
mainstream---is more a way of maintaining
individuality~\cite{our:seceder} than a relevant social
distinction. In our work we assume the mainstream can be regarded as a
collection or set of subcultures, kindred with the other subcultures,
though in reality of course by far the biggest. Media (though, by
definition, not the most used channels) has, of course, a role in
the spreading of more obscure subcultures too. A youth may adopt a
subculture only as a result of a media report. The real situation
there is a coupling from the youth to the media and back again. This
paper assumes the personal coupling is stronger when it comes to
identity formation and thus considers the spread through friends only.

Now we turn to the definition of our model. We use the framework of
graph theory and consider $N$ vertices connected pairwise by $M$
edges. The vertices represent the youths and the edges represent their
social ties. At time t the vertex i has a unique identity, or
subculture, $c(i,t)$. Time, in our simulation, is discrete. It is
represented by an integer number from 1 to $t_\mathrm{max}$ (we have
$t_\mathrm{max} = 12800$ for all runs of our simulations)
corresponding to the number of the iteration of the update of
$c(i,t)$. The central part of the updating procedure is the score
function $s_c(t,i)$ that is intended to represent the attractiveness
of subculture $c$ to person $i$ at time $t$. If the score-function
value of a subculture $c$ exceeds a threshold $T$ (which is a model
parameter) the individual will replace her/his current subcultural
identity by $c$. As mentioned, a high number of neighbours adopting
$c$ should make $c$ attractive. We make the simplest choice and let
$s_c(t,i)$ be a linear function of $n_i(c)$, the number of neighbours
of $i$ with the identity $c$. That the attractiveness of a subculture
decreases with time is implemented by making $s_c(t,i)$ proportional to
the age difference between $c$ and $c_i$, $i$'s current subculture,
divided by the current age of $c_i$. The last ingredient of the score
function is a normalisation factor, $k/k_i$, $k = 2M/N$ is the average
degree (number of neighbours of a vertex) and $k_i$ is $i$'s
degree. This factor is included to compensate for the varying degrees,
so that the same threshold value can be used for all vertices. To sum
up, the score function is
\begin{equation}\label{eq:score}
s_c(t,i) = \frac{k}{k_i} \: n_i(c)\: \frac{t(c)-t(c_i)}{t-t(c_i)} ,
\end{equation}
where $t(c)$ is the age of $c$ (note that if $t$ has no argument it
refers to the current simulation time). The iterations are as follows:
\begin{itemize}
\item For every vertex $i$ (chosen sequentially) calculate the score
  $s_c(t,i)$ of all subcultures $c$ for the individual $i$ at time $t$.
\item Go through the vertex set sequentially once again. If the score
  is higher than a threshold $T$ for some identity $c$, then $i$
  change its identity to $c$. If more than one subculture has a score
  above the threshold then the individual adopts the subculture with
  the highest score.
\item With a probability $R$ a new identity is assigned to a vertex. So,
  on average, $NR$ fads are introduced per time step.
\end{itemize}
An example of the propagation of subcultures on a small test graph can
be found in fig.~\ref{fig:ill}. The C++ source code can be found at
\url{http://www.tp.umu.se/forskning/networks/fads/}.

\section{Substrate networks}

\begin{figure}
  \resizebox*{0.95\linewidth}{!}{\includegraphics{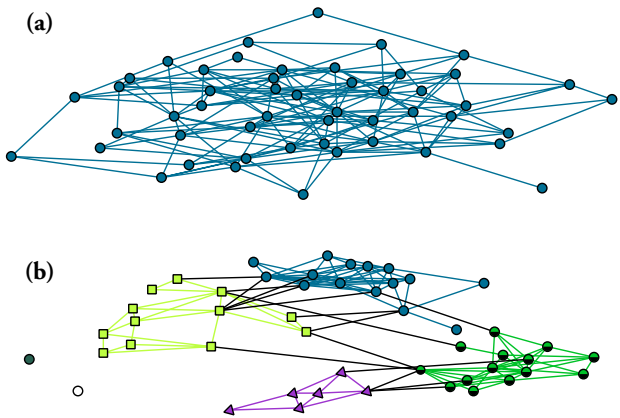}}
  \caption{Examples of network realisations of the ER (a) and the
    seceder (b) models. Both graphs have $N = 50$ vertices and $M = 150$
    edges, the seceder graph has the randomisation parameter value $r =
    0$. For the seceder model we indicate the communities, as
    identified by the algorithm of Ref.~\cite{mejn:fast}, by different
    symbols.
  }
  \label{fig:nwk}
\end{figure}

We will use two kinds of network models as underlying networks in our
study---one very basic model, the Erd\H{o}s-R\'{e}nyi (ER)
model~\cite{er:on,rap:cont}; and one modern model of acquaintance
networks, the networked seceder model~\cite{our:seceder}. We note
that, over the last decade, an abundance of network models have been
proposed, the reason we chose these two in particular is that they
represent two ends of the spectrum between generality and
specificity. The ER model is the simplest possible random graph
model---simple in the sense that it is maximally random with no
structural biases: One iteratively adds the edges between pairs of
vertices so that multiple edges and loops (self-edges) are avoided.

The seceder model is intended to generate networks with community
structure---densely connected subnetworks (communities) with
relatively few connections between the subnetworks. We will sketch how
this model works, for the exact details we refer to
Ref.~\cite{our:seceder}. One starts from an ER model network and
successively rewires (detach one side of an edge and attach it to some
other vertex) the edges according to the following approximate
rules: \textit{1.} Choose three vertices. \textit{2.} Select the one
$i$ of these with the highest eccentricity (maximal distance to any
other vertex~\cite{harary}). \textit{3.} Choose another vertex $j$ at
random and rewire $j$'s edges to $i$ and $i$'s
neighbourhood. \textit{4.} With a probability $r$ (the only parameter
of the seceder model) re-rewire one of $j$'s edges to a randomly
chosen vertex. With the parameter $r$ one can tune the randomness of
the model---with $r = 1$ the networks are of ER-type, with $r = 0$ the
community structure is most strongly pronounced.

The ER model networks are characterised by a Poissonian degree
distribution, a vanishing clustering (fraction of triangles) and no
pronounced community structure. The seceder model networks have an
exponentially decaying degree distribution, high clustering and strong
community structure. Two example networks are displayed in
fig.~\ref{fig:nwk}.

\section{Simulation results}

\begin{figure}
  \resizebox*{0.95\linewidth}{!}{\includegraphics{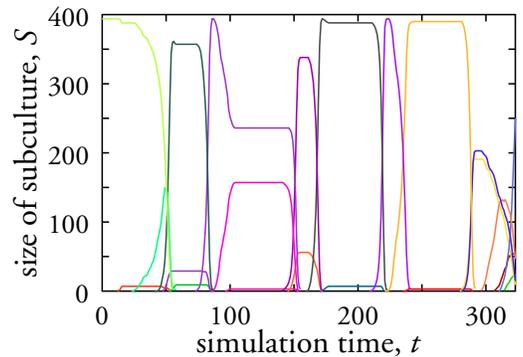}}
  \caption{Example of a run for ER networks with sizes $N = 400$, $M =
    800$, and threshold value $T = 0.8$. The plot shows the size of a few
    fads as a function of simulation time.}
  \label{fig:series}
\end{figure}

To get an impression of how subcultures are born, evolve and die, we
plot the time evolution of the number of adopters $S$ of different
fads. One typical run is seen in fig.~\ref{fig:series}. We observe
that, a subculture can most often be divided into a growth stage, a
quasi-stationary stage and a final decline. We note that this
distinction of three stages is rather
common~\cite{thornton,damrell}. The shape and slope of the growth and
decline, the life-length and the maximal size are all distinct for
different subcultures. Some fads have more complex time evolution with
more than one quasi-stationary state. The time evolution for the
seceder model is more complex than for the ER model networks seen in
fig.~\ref{fig:series}. We note that two leading time-scales of this
model are the life-length of subcultures and the mean time between
introductions of new subcultures. If we have a scale separation (i.e.\
that, either the life-length is much larger than the introduction
time, or vice versa), then the model can be simplified. In the limit
of short time between the introductions of new subcultures, every
individual will have a new subculture with each time step. In the opposite
limit the model essentially reduces to the threshold model of
Watts~\cite{watts:fad}. But in the real world these two time scales
are not separable---some subcultures may flare up and vanish within
the life-span of others. Our model is thus relevant in this
intermediate region and our choice of $R = 1/6400$ (that we use
throughout the simulations) as seen in fig.~\ref{fig:series}, ensures
this. Henceforth, we will study quantities averaged over $\sim 250$
realisations of the graph models, this means that there will be $\sim
250\, t_\mathrm{max}\, R\, N = 500N$ fads introduced per point in
parameter space.

\begin{figure*}
  \resizebox*{0.9\linewidth}{!}{\includegraphics{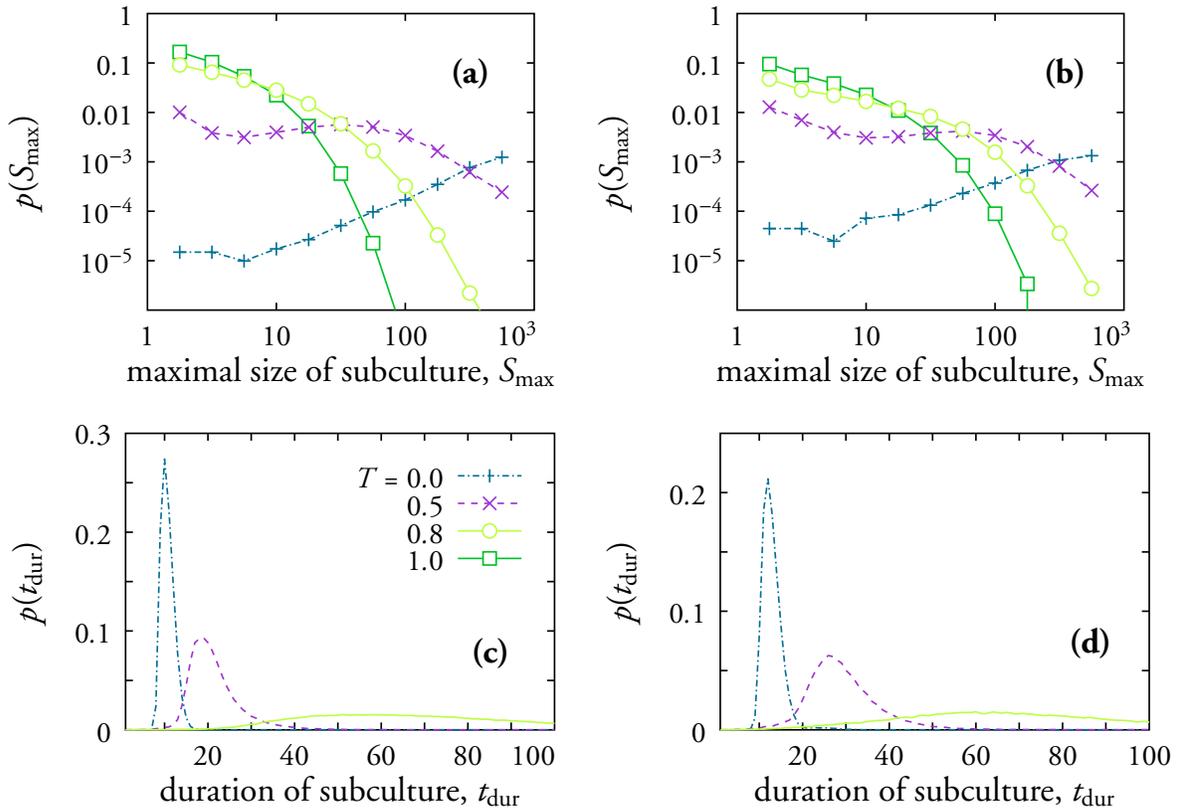}}
  \caption{The probability density function of the largest size of a
    subculture $S_\mathrm{max}$ for the ER (a) and seceder (b) models,
    and the probability density function of the life-length
    ($t_\mathrm{dur}$) of subcultures for ER (a) and seceder (b)
    models. The network sizes are $N = 1000$ and $M = 2000$, for the
    seceder model we have the randomness-parameter value $r =
    0.1$. Errorbars are smaller than the symbol sizes.}
  \label{fig:nt}
\end{figure*}

In fig.~\ref{fig:nt}(a) and (b) we display the distribution plots of
the maximal size of the subculture $S_\mathrm{max}$. We see that
different threshold values give qualitatively different distributions
of $S_\mathrm{max}$. For small threshold values $p(S_\mathrm{max})$
grows with $S_\mathrm{max}$. This means that most subcultures will, at
their peak, affect most of the population. For small $T$-values, or $T
= 0$, $S_\mathrm{max}$ is decaying sharply (with an exponential
tail). For intermediate values of $T$ there will be a very broad
distribution of the maximal subculture-size---a situation where some
subcultures grow to involve almost every individual, while many others
die out without gaining many followers. Note that such a
broad distribution is not a trivial result of the introduction
dynamics---a new subculture, in our model, enters the population
according to the sharp Poisson-distribution. The two different
underlying networks---the ER model of fig.~\ref{fig:nt}(a) and seceder
model of fig.~\ref{fig:nt}(b)---do not cause any qualitatively
different behaviour. (This conclusion---that the result is
qualitatively independent of the underlying network structure---is
confirmed in preliminary tests on different network models.) There
seceder model curves are slightly flatter indicating that there is a
larger probability for small subcultures for low $T$-values and a
larger chance for population wide subcultures if $T$ is high. A
natural explanation for this lies in the community structure of the
seceder model. Since these communities are highly interlinked, the
probability that a vertex is neighbour to more than one vertex of
the same subcultural identity is higher. Therefore, a new subculture
can more easily overcome the threshold and spread within a community. For
the same reason it is harder for a subculture to spread between
communities. The blurring of the curves of fig.~\ref{fig:nt}(b)
(compared with those in fig.~\ref{fig:nt}(a)) is thus to some extent
due to the community structure. This conclusion is consistent with the
plots of the total life time of subcultures $t_\mathrm{dur}$ shown in
fig.~\ref{fig:nt}(c) and (d). The seceder model curves
(fig.~\ref{fig:nt}(c)) shows a broader distribution of life-lengths
than the ER model (fig.~\ref{fig:nt}(d)), which can be explained by
that the fate of a subculture is, to some extent, determined by the
community in which it has started. Another observation regarding the
$t_\mathrm{dur}$-distribution is that the peaks come later on the
seceder model. This can be explained by the higher clustering
(fraction of complete triangles with respect to connected
vertex-triples) of the seceder model. A high clustering gives a
correction to the number of individuals $m$ step away from a given
vertex~\cite{mejn:clumodel}---this number is smaller for a network of
high clustering, which means that the number individuals reached (in a
person-to-person spreading process) after a certain time is also
decreases with the clustering.

\begin{figure}
  \resizebox*{0.95\linewidth}{!}{\includegraphics{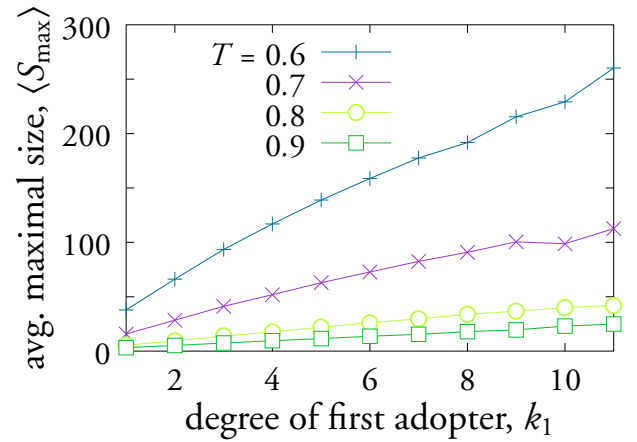}}
  \caption{The average maximal size $\langle S_\mathrm{max}\rangle$ of
    a subculture as a function of the degree of the first adopter
    $k_1$. The network is of ER type with sizes $N = 1600$ and $M =
    3200$. Errorbars are of the order of the symbol size.}
  \label{fig:ndeg}
\end{figure}

So the broad distribution of subculture sizes can be explained by the
model, but can anything be said how a large subculture is initiated?
Intuitively one expects that if the first adopter has higher degree
then the subculture has a higher chance of being a population-wide
fad. In fig.~\ref{fig:ndeg} we plot the average peak-size $\langle
S_\mathrm{max}\rangle$ as a function of the degree of the first
adopter $k_1$. As expected $\langle S_\mathrm{max}\rangle (k_1)$ is an
(sublinearly) increasing function. If the threshold is lower the
influence of high-degree vertices increases. This is an effect of the
larger average peak sizes of lower $T$-values. (For higher $T$-values more
subcultures coexist, so the peak sizes are smaller but the life-length
longer.)

\begin{figure}
  \resizebox*{0.95\linewidth}{!}{\includegraphics{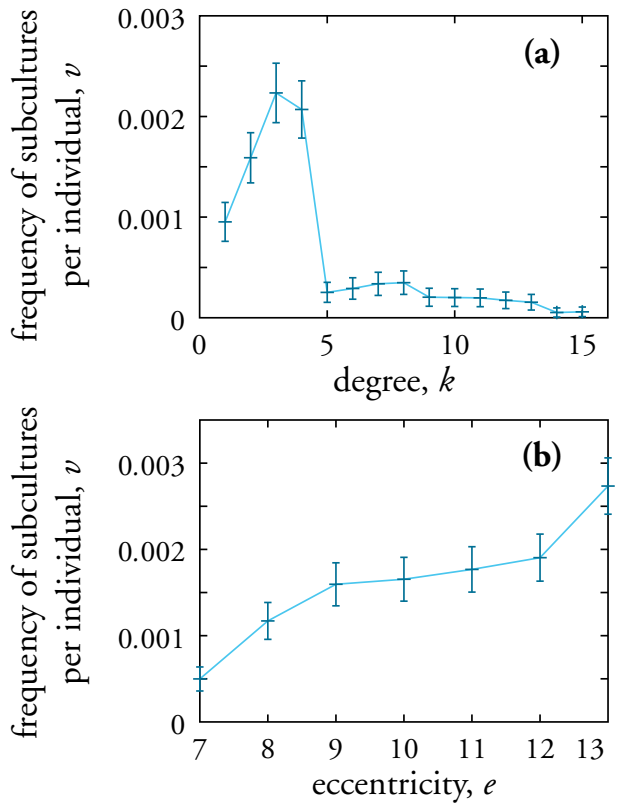}}
  \caption{The distribution of the frequency of subcultures per
    individual $v$ as a function of degree (a) and eccentricity (b). The
    underlying networks are of ER type with $N = 1000$ vertices $M = 2000$
    edges and threshold value $T = 1.0$.}
  \label{fig:acti}
\end{figure}

In addition to the phenomenon that subcultures can be central for the
formation of a youth's personal identity, some adolescents shift such
characteristics more frequently than others. This can of course be
modelled by a varying threshold, but may also be an effect of the
structure of the social network. In fig.~\ref{fig:acti}(a) we plot the
frequency $v$ of subcultures per individual, i.e.\ the number of
subcultures that can be expected to be adopted during one time
step. Since only one new subculture can be adapted per time step $v$
is also the probability that an individual will adopt a new identity a
particular time step. We see a sharp peak of $v$ for small
$k$-values. Low-degree vertices are thus, by this property alone, more
prone to changing their identity. In fig.~\ref{fig:acti}(b) we plot
$v$ as a function of the eccentricity $e$ (the maximal distance to any
other vertex in the network) of the vertex. We observe that $v$ is an
increasing function of $e$. To summarise, individuals that have few
acquaintances and are peripheral in the social networks changes
identity more often. One observation in favour of this finding is
Thornton's study~\cite{thornton} of British club cultures where many
adopters change as soon as a specific style has been adopted by the
mainstream. Apart from this we have not found any empirical
observations of this phenomenon. Even if the non-constant $v$ of
fig.~\ref{fig:acti} would be dismissed as an artifact of our model, it
still exemplifies that it is not only individual characteristics that
create the social network---one's identity may also be formed by the
social network structure.

\section{Summary and conclusions}

We have presented a model for how subcultures spread in a population
of adolescents. Using five main assumptions of a youth's response to
the subcultures of others to whom they are socially close, we
construct dynamical rules
for subculture diffusion. This dynamical model is then put on networks
intended to represent acquaintance ties. The model is sketchy, but
contains, we believe, many of the important mechanisms for the
evolution of the youth subcultures. Even if each individual has a very
complex rationale for her, or his, response to the social environment,
the fact that individuals, on average, respond in certain ways makes
it possible for only a few mechanisms to govern the system-wide
properties~\cite{schelling}. One problem with the particular question
we address, is the lack of quantitative empirical data. Unlike the
related issue of memberships in voluntary
organisations~\cite{liljeros:phd} an adolescent is not required to
register in any way. One can, as we do, compare qualitative model
behaviour with qualitative observations. But for future studies it
should be possible to perform quantitative studies; either directly by
longitudinal interview surveys, or indirectly by e.g.\ measuring the
frequency of related key words in the press.

One characteristic our model shares with real world observations is
that, for certain parameter values, a few subcultures have a much
larger staying-power than the average, whereas most subcultures die
out very soon. The punk music scene of the 1970's~\cite{rtr} is an
example of a long-lived and large subculture. The other end, that of
short-lived and small subcultures, is less well-defined: one can
imagine peculiar habits of a circle of friends, maybe even one person,
to define a subculture. Such short-lived fads can be great
in number and still remain largely unheard of just because the total
number of practitioners is small. We find that the maximal size of the
networks is strongly correlated with the degree of the first
adopter. Another outcome of our model is that fringe groups change
identity more frequently than central and well-connected
vertices. This may or may not reflect a real situation, but it serves
as an example of how traits of individuals can result from the social
structure, as well as vice versa. The qualitative behaviour of our
dynamical rules is the same for our two network models. This means
that the result is stable for moderate changes of the underlying
network topology, and therefore is likely to remain valid for large
ranges of real social networks.

Many studies separate subcultures and the commercialised mainstream
youth cultures. It has been argued that this dichotomy, rather than
being an accurate social description, stems from members of smaller
subcultures and their need to profile themselves as different opposed
to a larger mainstream culture~\cite{thornton}. In our model, we would
like to interpret the largest subcultures as forming the
mainstream. Naturally, these subcultures, rather than smaller
subcultures, tend to be the focus of commercial interests.

Many extension and versions of this and similar threshold models are
possible. We anticipate future work proposing, just as we (in this
paper) and e.g.\ Ref.~\cite{upal} do, fine tuned models for
information dissemination on more specific social systems than more
general sociodynamic models of e.g.\ information diffusion and opinion
formation (see
e.g.\ Refs.~\cite{sznajd,bennaim:compromise,dodds:contagion} and
references therein).

\begin{acknowledgments}
Thanks are due to Elizabeth Leicht for critically reading the
manuscript and to Beom Jun Kim for comments and partial financial
support.
\end{acknowledgments}

\end{document}